\documentclass[jkps,twocolumn,fleqn,showpacs,showkeys]{revtex4}   

\usepackage{graphicx}
\usepackage{amssymb}
\usepackage{amsmath}
\usepackage{bm}
\usepackage{times}

\newcommand{\msun}{M_\odot}

\newcommand{\be}{\begin{equation}}
\newcommand{\ee}{\end{equation}}
\newcommand{\bea}{\begin{eqnarray}}
\newcommand{\eea}{\end{eqnarray}}

\newcommand{\etal}{{\it et al.}}

\begin{document}
\setcounter{page}{1}

\title[]{Systematic Bias Due to Nonspinning Template Waveforms in the Gravitational Wave Parameter Estimation for Aligned-Spin Binary Black Holes}

\author{Hee-Suk \surname{Cho}}
\email{chohs1439@pusan.ac.kr}
\affiliation{Department of Physics, Pusan National University, Busan 46241, Korea}

\date[]{}

\begin{abstract}
We study the parameter estimation of gravitational waves for aligned-spin binary black hole (BBH) signals
and assess the impact of bias that can be produced by using nonspinning  template waveforms. 
We employ simple methods to calculate the statistical uncertainty from an overlap distribution.
For the fiducial waveform model, we use a phenomenological model, which is designed to generate the gravitational waveforms emitted from merging BBH systems.
We show that the mass parameters recovered by using nonspinning waveform templates can be significantly biased  from the true values  of aligned-spin signals.
By comparing the systematic bias with the statistical uncertainty, 
we examine the validity of nonspinning templates for the parameter estimation of aligned-spin BBHs.
\end{abstract}



\maketitle

\section{INTRODUCTION}
Since the prediction of gravitational radiation from Einstein's general theory of relativity, no direct detection has been made for about 100 years because the gravitational wave (GW) strain is extremely weak.
At last, however,  two GW signals, named as GW150914 and GW151226, were recently detected by the two LIGO detectors \cite{GW1,GW2},
and future observing runs of the advanced detector network \cite{ALIGO,AVirgo,KAGRA} are expected to yield more binary black hole (BBH) merger signals \cite{Aba10,Dom14,Abb16,Abb16b}.
Detailed analyses of the two GWs showed that both signals were emitted from merging BBHs \cite{GW1PE1,GW1PE2,GW2}.
The masses of the two binaries were found to be $\sim 65 \msun$ and $\sim 22 \msun$ for GW150914 and GW151226, respectively.  
In particular, the masses of the two component BHs  of GW150914 were estimated as $\sim 36 \msun$ and $\sim 29 \msun$, 
and those are believed to be the heaviest stellar-mass BHs known to date.

In GW data analyses for ground-based detectors, a GW event is identified in the search pipeline
by using the matched filtering method \cite{All12}. 
Once a detection is made in the search,
the parameter estimation pipeline seeks the physical properties of the GW source, such as mass and spin \cite{Aas13b}.
Generally, this procedure takes a very long time, up to several
months depending on the waveform model used in the analysis \cite{Cor11,Vei12,Osh14a,Osh14b}.
The result of the parameter estimation is given as
probability density functions (PDFs) for all parameters and their correlations. 
The statistical uncertainty given in the PDF indicates how accurately
the parameter can be measured by the analysis,
and this measurement error depends on the strength of the GW signal.
In principle, the statistical uncertainty is inversely proportional to the signal-to-noise ratio (SNR) \cite{Fin92,Cut94}.  
On the other hand, the result of the parameter estimation fundamentally includes another type of error,
the systematic bias \cite{Cut07}. Unlike the statistical error, the systematic error relies on how exactly the
wave function mimics real gravitational waveforms.
For example,  if a template waveform model neglects some physical parameters to simplify the wave function,
this model cannot exactly describe real GW signals, and the parameter estimation analysis will give a wrong result.
Typically, the validity of a waveform model used for the parameter estimation can be examined by comparing the systematic bias with the statistical  uncertainty \cite{Fav14,Cho15d}.

In this work,  we investigate the statistical and the systematic errors in the parameter estimation for aligned-spin BBH signals.
We adopt a simple approach to calculate the confidence region directly from an overlap distribution \cite{Bai13}
and use the effective Fisher matrix method to determine the statistical uncertainty analytically \cite{Cho13,Cho14}.
We calculate the bias that is produced by using nonspinning templates, and
by comparing the biases with the statistical errors,  
we examine the validity of the templates in the parameter estimation for aligned-spin BBH signals.
As our fiducial gravitational waveform model, we adopt a phenomenological model, which is an analytic approach designed to
generate the gravitational waveforms emitted from aligned-spin BBH systems \cite{Kha16}.


\section{statistical error and systematic bias in the parameter estimation} 
The match between a detector data stream $x(t)$ and a template $h(t)$ can be defined as
\be \label{eq.match}
\langle x | h \rangle = 4 {\rm Re} \int_{f_{\rm low}}^{\infty}  \frac{\tilde{x}(f)\tilde{h}^*(f)}{S_n(f)} df,
\ee
where $S_n(f)$ is the power spectral density (PSD) of the detector noise, and
$f_{\rm low}$ is the low-frequency cutoff that depends on the detector's frequency band.
The tilde denotes the Fourier transform of the time-domain waveform.
In this work, we use the zero-detuned, high-power noise PSD, which is one of the designed sensitivities for Advanced LIGO \cite{apsd}, 
and assume $f_{\rm low}=10$ Hz.
For aligned-spin BBH systems in circular orbits, the wave function can be described by  using 11 parameters:
those are five extrinsic parameters (luminosity distance of the binary, two angles defining the sky position of the binary with respect to the detector, orbital inclination, and wave polarization), two arbitrary constants (coalescence time $t_c$ and coalescence phase $\phi_c$),
and  four intrinsic parameters (two masses $m_1, m_2$ and two spins $\chi_1, \chi_2$).
However, the extrinsic parameters only scale the wave amplitude and do not affect our analysis. 
In addition, the contribution of the two arbitrary constants can be easily removed from our analysis 
by using certain analytic techniques \cite{All12}.
Therefore, we consider only four parameters, $m_1, m_2, \chi_1$ and $\chi_2$, in our match calculation.

We define the overlap $P$ by the match between  a normalized signal $\hat{h}_s$ and a normalized template $\hat{h}_t$ maximized over $t_c$ and $\phi_c$:
\be\label{eq.overlap}
P =  \max_{t_c,\phi_c}\langle \hat{h}_s | \hat{h}_t \rangle. 
\ee
If $\hat{h}_s = \hat{h}_t$, we obtain $P=1$.
Since we take into account nonspinning templates in this work, the overlap function is distributed on
the two-dimensional surface as
\be\label{eq.2d overlap}
P(\lambda) =  \max_{t_c,\phi_c}\langle \hat{h}_s(\lambda_0) | \hat{h}_t(\lambda) \rangle, 
\ee
where $\lambda_0$ denotes a set of the true parameters of the signal, 
and $\lambda$ varies in the $m_1-m_2$ plane.
In this equation, while the signal wave function has a spin parameter, the templates are nonspinning waveforms.
Therefore, if the signal has a non-zero spin value, the mass parameters recovered by using the nonspinning templates can be  
systematically biased from the true parameters.
The bias can be determined by the distance from the true value $\lambda_0$ to the recovered value $\lambda^{\rm rec}$:
\be\label{eq.bias}
b =   \lambda^{\rm rec} - \lambda_{0}.
\ee
In the parameter estimation, the bias corresponds to the systematic error, and the magnitude of the bias is associated with the accuracy of the template waveform model.
The impact of bias on the parameter estimation can be measured by comparing that with the statistical uncertainty.
Below, we briefly describe how the statistical uncertainty can be calculated from the overlap distribution.

The PDF of the parameter estimation is given by the likelihood function ($L$),
and the likelihood is related to the overlap  as \cite{Cho13}
\be \label{eq.LvsP}
\ln L(\lambda) =-\rho^2  (1-P(\lambda)),
\ee
where $\rho$ is the SNR.
From this relation, one can infer that the confidence region of the PDF 
is directly connected with a certain  region in the overlap surface.
Baird {\it et al.}~\cite{Bai13} showed that if the likelihood has a Gaussian distribution,
the confidence region is approximately consistent with the overlap region that is surrounded by  the iso-match contour (IMC) determined by
\be \label{eq.IMC}
P=1-{\chi^2_k(1-p) \over 2 \rho^2 },
\ee
where $\chi^2_k(1-p)$ is the chi-square value for which probability of obtaining that value or larger  is $1-p$, and $k$ denotes the degree of freedom, which is given by the number of parameters.
In this work, since we explore two-dimensional mass parameter space (i.e., $k=2$), the one-sigma confidence region (i.e., $p=0.682$) with $\rho=20$ is given by the contour $P=0.99714$.

Next, in order to calculate the statistical uncertainty and the correlation coefficient, we employ the Fisher matrix formalism
(for more details refer to \cite{Val08} and references therein).
The Fisher matrix can be calculated by using the overlap as \cite{Cho13,Cho14,Jar94}
\be \label{eq.eFM1}
\Gamma_{ij}=\bigg\langle {\partial \tilde{h} \over \partial \lambda_i} \bigg | {\partial \tilde{h} \over \partial \lambda_j} \bigg \rangle\bigg|_{\lambda=\lambda_0}
\simeq -\rho^2 \frac{\partial^2 P(\lambda)}{\partial \lambda_i \partial \lambda_j}\bigg|_{\lambda=\lambda_0}.
\ee
If the overlap surface has a quadratic shape,
one can define a quadratic function $F$ that fits the overlap $P$.
Using this function, Cho \etal  \cite{Cho13} defined the effective Fisher matrix as
\be \label{eq.eFM2}
(\Gamma_{ij})_{\rm eff}= -\rho^2 \frac{\partial^2 F(\lambda)}{\partial \lambda_i \partial \lambda_j}\bigg|_{\lambda=\lambda_0}.
\ee
The inverse of $\Gamma_{ij}$ represents the covariance matrix of the parameter errors, 
and the statistical error ($\sigma_i$) of each parameter and the correlation coefficient ($c_{ij}$) between two parameters are determined by
\be \label{eq.sigma}
\sigma_i=\sqrt{(\Gamma^{-1})_{ii}},  \ \ \  {\cal C}_{ij}={(\Gamma^{-1})_{ij} \over \sqrt{(\Gamma^{-1})_{ii} (\Gamma^{-1})_{jj}}}.
\ee
On the other hand, if the waveform model is expressed by using an analytic function in the Fourier domain, such as in post-Newtonian models,
the Fisher matrix can be directly computed from the wave function as in Eq. (\ref{eq.eFM1}) \cite{Cho15a,Cho15b,Cho16}. 
However, the functional form of the phenomenological waveform model is much more complicated compared to the post-Newtonian models,
so we adopt the effective Fisher matrix method in this work.

\section{phenomenological waveform model} 
The phenomenological waveform model is designed to produce the gravitational waveforms
emitted from merging BBH systems. So far, various versions have been developed for nonspinning \cite{Aji07a,Aji08a,Aji08b}, aligned-spin \cite{Aji11b,San10,Kha16}, and precessing-spin \cite{Han14} systems. 
In this work, we use the most recent phenomenological model, the so-called PhenomD \cite{Kha16}.
The phenomenological wave function defined in  the Fourier domain has the form
\be \label{eq.phenom}
\tilde{h}(f)=A_{\rm eff}(f) e^{i \Psi_{\rm eff}(f)},
\ee
where $A_{\rm eff}$ and $\Psi_{\rm eff}$ are the effective amplitude and the effective phase, and those are modeled separately 
and parameterized by using the mass and the spin parameters \cite{Cho15c}.

On the other hand, for aligned-spin systems, the two spins are strongly correlated \cite{San10,Aji11a, Pur13,Nie13,Pur16b}.
Thus,  treating the effect of aligned-spins with a single spin parameter defined by
\be \label{eq.effective spin}
\chi \equiv \frac{m_1 \chi_1+ m_2 \chi_2}{M},
\ee
is more efficient, where $M$ is the total mass.
In the PhenomD wave function, the value of $\chi$ can be determined simply by choosing  $\chi_1=\chi_2=\chi$.
In this work, therefore, the signal waveforms are given by $h_s=h(m_1, m_2, \chi)=h_{\rm PhenomD}(m_1, m_2, \chi, \chi)$, and
the template waveforms are given by $h_t=h(m_1,m_2)=h_{\rm PhenomD}(m_1, m_2, 0, 0)$ \cite{Cho16b}.
On the other hand, we use the chirp mass $M_c\equiv (m_1 m_2)^{3/5}/  (m_1+m_2)^{1/5}$ and the symmetric mass ratio $\eta \equiv m_1 m_2 / (m_1 + m_2)^2$ because those are more efficient than $m_1, m_2$ in our analysis \cite{Cho15a}.


\section{Result}


\subsection{Statistical Error}
\begin{figure}[t]
\begin{center}
\includegraphics[width=\columnwidth]{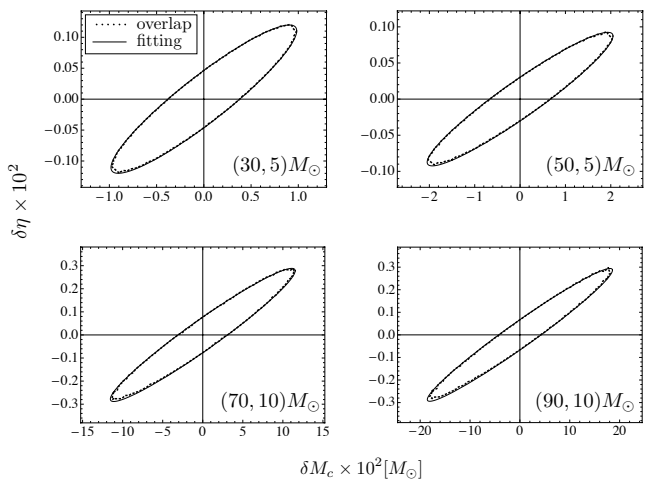}
\caption{\label{fig.overlap}One-sigma confidence region with $\rho=20$ ($P=0.99714$), and its quadratic fitting ellipse.}
\end{center}
\end{figure}

We choose five BBH signals with masses of  $(30, 5) \msun$, $(50, 5) \msun$, $(50, 10) \msun$, $(70, 10) \msun$, and $(90, 10) \msun$.  Using Eq. (\ref{eq.2d overlap}), we calculate the overlap surfaces, 
where we have considered the signals with $\chi=0$ to obtain unbiased overlap surfaces. 
From the IMC approach given in Eq. (\ref{eq.IMC}), the one-sigma confidence region with $\rho=20$ can be given by the contour $P=0.99714$. 
On the other hand, the IMC method was established under the assumption that the PDF obeys a Gaussian distribution \cite{Bai13}.
Thus, the accuracy of this method  relies on the Gaussianity of the likelihood \cite{Has15},
and from Eq. (\ref{eq.LvsP}), a Gaussian likelihood corresponds to a quadratic overlap in the region given by the SNR.
Therefore, the confidence region obtained by using the IMC method  can be reliable if the overlap surface has a quadratic shape in that region.
To verify this,  in Fig. \ref{fig.overlap}, we show the confidence region  (i.e., the contour of $P=0.99714$) and its fitting ellipse determined by using the quadratic fitting function $F$.
One can see that the overlap contours almost exactly overlap their fitting ellipses.
On the other hand, Berry \etal~\cite{Ber15} investigated the effect of  realistic non-Gaussian noises  on the parameter estimation and 
found that  the character of the noise made negligible difference to the PDFs.

Next, we calculate the effective Fisher matrix by inserting the fitting function $F$ into Eq. (\ref{eq.eFM2})
and compute the statistical uncertainty and the correlation coefficient by using Eq. (\ref{eq.sigma}).
In Table~\ref{tab.1},  we summarize the results for our BBH signals.
Here, from the relation $M=M_c\eta^{-3/5}$, $\sigma_M$ can be computed in the following way:
\bea
\sigma_M^2 &=&\left(\frac{\partial M}{\partial M_c}\right)^2 \sigma_{M_c}^2+\left(\frac{\partial M}{\partial \eta}\right)^2 \sigma_{\eta}^2  \\ \nonumber
&&+ 2 {\cal C}_{M_c \eta}\left(\frac{\partial M}{\partial M_c}\right)\left(\frac{\partial M}{\partial \eta}\right) \sigma_{M_c} \sigma_{\eta}.
\eea
We find that the fractional statistical errors ($\sigma_{\lambda}/\lambda$) for the mass parameters increase as the total mass increases.
We also find that the correlation coefficient between $M_c$ and $\eta$ increases with increasing total mass,
and this trend can be inferred from the ellipses in Fig. \ref{fig.overlap}.

\begin{table}[t]
\caption{\label{tab.1}{Statistical uncertainties and correlation coefficients calculated by using the effective Fisher matrix method for BBH signals with $\rho=20$.}}
\begin{ruledtabular}
\begin{tabular}{c  ccccc }
$m_1 [\msun], \ m_2 [\msun]$  & 30, 5  & 50, 5 &50, 10 & 70, 10 &90, 10 \\
   \hline
 $\sigma_{M_c}/M_c \times 10^2$	    &0.065                &0.110        &0.250                        &0.359   &0.520\\
 $\sigma_{\eta}/\eta  \times 10^2$	      &0.647                  &0.739      &1.469                      &1.741  &      2.159\\
 $ {\cal C}_{M_c \eta}$	      &0.921                  &0.944      &0.952                     &0.963  &      0.973\\
 $\sigma_M/M \times 10^2$                &0.329            &0.341           &0.648                    &0.705&       0.798
  \end{tabular}
 \end{ruledtabular}
\end{table}


\subsection{Impact of the Systematic Bias}
\begin{figure}[t]
         \includegraphics[width=\columnwidth]{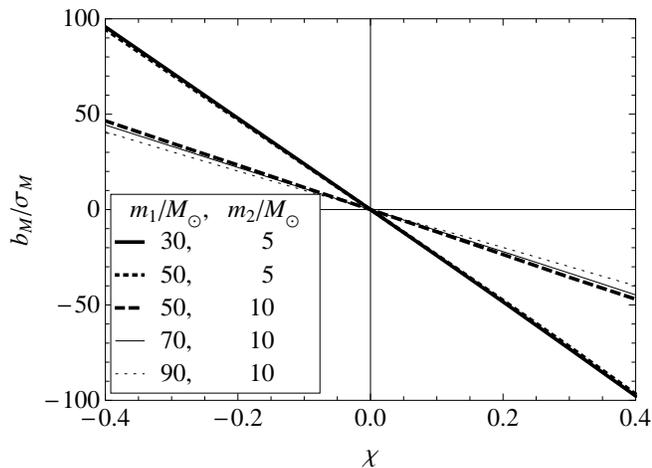}
    \caption{\label{fig.fracbias-M} Comparison between the systematic bias ($b$) and the statistical uncertainty ($\sigma$) for the parameter $M$.}
\end{figure}

We now consider the spin for our signals and calculate the overlaps by using Eq. (\ref{eq.2d overlap}).
Since our templates do not have spins, the resultant overlap surfaces are biased,
and the bias can be measured by using Eq. (\ref{eq.bias}).
Basically, the templates cannot be reliable for the parameter estimation if the recovered parameter is significantly biased
from the true parameter. 
To examine the validity of nonspinning templates for aligned-spin signals,
we compare the systematic bias with the statistical uncertainty for the parameter $M$.
The result is given in Fig. \ref{fig.fracbias-M}, where we have adopted the statistical uncertainties given in Table \ref{tab.1}.
We find that the systematic errors are much larger than the statistical errors, even with a very small value of the spin.
In addition, the ratio $b/\sigma$ can be larger for  higher SNRs because the statistical error is inversely proportional to the SNR.
Therefore, we conclude that nonspinning templates are unsuitable for the parameter estimation for aligned-spin BBH signals.
On the other hand, one can see a linear dependence of the ratio $b/\sigma$ on the spin in the range of $-0.4 \leq \chi \leq 0.4$ for all signals.
Interestingly, depending on the mass of the lighter BH rather than the total mass, the results in this figure show clearly separated populations.
This is because the statistical error ($\sigma_M/M$)  depends mainly on the  mass of the lighter BH, as shown in Table \ref{tab.1},
while the biases ($b_M/M$) are mostly consistent independently of the masses. 

\section{Summary and Conclusion}\label{sec5}
We studied the reliability of nonspinning templates in the GW parameter estimation for aligned-spin BBH signals.
Assuming several signals with various masses,  we calculated the statistical uncertainty
by using the IMC and the effective Fisher matrix methods.
We also calculated the systematic biases induced by nonspinning templates for the aligned-spin signals
and compared those with the statistical uncertainties to examine the impact of bias on the parameter estimation.
We found that  the total mass recovered by using the nonspinning templates could be significantly biased 
from the true value of the aligned-spin signal.
The bias can exceed the statistical error even with a very small value of the spin.
Therefore, nonspinning templates are unsuitable for the parameter estimation for aligned-spin BBH signals.


%

\section*{ACKNOWLEDGMENTS}
This work was supported by the National Research Foundation of Korea (NRF) grant funded by the Korea government (Ministry of Science, ICT \& Future Planning) (No. 2016R1C1B2010064 and No. 2015R1A2A2A01004238).
This work used the computing resources at the KISTI Global Science Experimental Data Hub Center (GSDC).
%
%
%

\end{document}